\documentclass[12pt]{article}
\usepackage{amsfonts,amsmath,amsxtra}
\newcommand{\beq}[1]{\begin{equation}\label{#1}}
\newcommand\eeq {\end{equation}}
\newcommand\bqa {\begin{eqnarray}}
\newcommand\eqa {\end{eqnarray}}
\newcommand\pr {\partial}

\newcommand\nn {\nonumber}

\newcommand{\bear}{\begin{array}}
\newcommand{\enar}{\end{array}}

\newcommand{\R}{\mathbb{R}}

\newcommand{\A}{\mathbb{A}}
\newcommand{\Z}{\mathbb{Z}}

\begin{document}

\def\t{\theta}
\def\T{\Theta}
\def\w{\omega}
\def\ov{\overline}
\def\a{\alpha}
\def\b{\beta}
\def\g{\gamma}
\def\s{\sigma}
\def\l{\lambda}
\def\wt{\widetilde}
\begin{flushright}
\hfill{hep-th/0105228}\\ \hfill{ITEP-TH-19/01 \,} \\ \hfill{YCTP -
SS5-01 \,}
\end{flushright}

\vspace{10mm}

\centerline{\Large \bf On Unification of RR Couplings} \vspace{10
mm} \centerline{{\bf Emil T. Akhmedov}\footnote{Permanent Address:
Institute of Theoretical and Experimental Physics, B.
Cheremushkinskaya 25, 117259 Moscow, Russia.},}
\centerline{Department of Physics and Astronomy,}
\centerline{University of British Columbia,}
\centerline{Vancouver, British Columbia, Canada V6T 1Z1}
\vspace{10mm} \centerline{{\bf Anton A. Gerasimov}}
\centerline{Institute of Theoretical and Experimental Physics,}
\centerline{B. Cheremushkinskaya 25, 117259 Moscow, Russia}
\vspace{5mm} \centerline{and} \vspace{5mm} \centerline{{\bf Samson
L. Shatashvili} \footnote{On leave of absence from St. Petersburg
Branch of Steklov Mathematical Institute, Fontanka, St.
Petersburg, Russia.}} \centerline{ Yale University, Department of
Physics,} \centerline{P. O. Box 208120, New Haven, CT 06520-8120}
\vspace{10mm}

\centerline{\bf Abstract} \vspace{0.5 cm} We consider the
couplings of RR fields with open string sector for
$Dp$-${\overline{Dp}}$ backgrounds of various $p$. The proposed
approach, based on the approximation of the open string algebra by
the algebra of differential operators, provides the unified
description of these couplings and their interrelations.

\vspace{10 mm}
\section{Introduction}
The process of brane annihilation \cite{Sen} leads to the unified
description of the backgrounds with various $D$-brane
configurations \cite{WitK}. However, despite the recent progress
in understanding  the tachyon low energy action and the related
qualitative phenomena \cite{GS1,KMM,KMMSup, GS2}, the picture of
the dynamics of the $D$-brane formation/annihilation is still not
very clear. One of the most important points seems to be the
interrelation of the open and closed string sectors in this
process, which is one of the main mysteries of string theory. The
satisfactory understanding of this relation most likely will
provide the clue regarding the symmetries of string theory
\cite{GS2}.

In this paper we reconsider the couplings of the RR gauge fields of
the closed string sector with the fields of the open string sector.
These couplings \cite{StCS} are especially simple due to their
anomalous origin \cite{Moore1} and thus provide a suitable
framework for the discussion of the various open string
backgrounds. These couplings were extensively studied (see e. g.
\cite{BCR,FW,Supercon,KeWi,Hori,KrLar,TaTeTa,oz,
Minasone,Minastwo,SWit}). Our ultimate goal will be the unified
description of these couplings for various $D-\overline{D}$
backgrounds\footnote{We were informed by G. Moore that some related
studies have been undertaken by him in collaboration with
Duiliu-Emanuel Diaconescu.}. The picture that emerges could be
considered as an explicit realization of some proposals from
\cite{WitN} (some related remarks have been made in \cite{oz}).

In an
attempt to provide the unified description of the RR gauge field
couplings for various background brane configurations it is
natural to consider the off-shell interpolation of these
backgrounds. The connection with K-theory uncovered in
\cite{Moore,WitK} leads to the description of the RR-gauge field
couplings in terms of the superconnection \cite{Quillen}. Thus a
natural framework for the universal description of RR couplings
would be in terms of a superconnection in some universal
(infinite) dimensional bundle $\mathcal{E}$ over space-time $M$
which provides interpolation between various brane/anti-brane
configurations. This infinite-dimensional bundle $\mathcal{E}$
should be naturally connected with the geometry of $M$ and thus it
is not very surprising to find the "tautological" bundle with
infinite dimensional fiber isomorphic to space of functions on the
base manifolds. This bundle has a rich structure connected with
the action of the differential operators in the fiber. This is in
perfect agreement with the appearance of the differential
operators in the explicit description of the elements of the
K-homolgy groups.

The appearance of the differential operators is
also natural from string field theory point of view. Configuration
space of the open string theory is roughly given by the maps of
the interval into space-time. If we approximate the strings by
straight lines the configuration space reduces to space of the
pairs of the points (ends of the intervals) in space-time.
Therefore the functionals on configuration space become the
functions of two points and could be interpreted as the kernels of
the integral operators. The expansion around diagonal (image of
the interval is a point) leads naturally to the differential
operators.

In this note we discuss very briefly the approximation of the open
string algebra by the differential operators. The more systematic
exposition will is given in \cite{First}. Here we describe
explicit interpolation of the various RR gauge field couplings
using the formalism of superconnections. For simplicity we always
consider the case of the flat space-time. The generalizations to
curved manifolds are rather obvious. In the process of describing
our proposal we will focus on the local properties.

The organization of this paper is as follows. In section 2 we
consider the interpolation between the RR gauge field couplings
with open strings for various brane/anti-brane backgrounds in
terms of superconnections acting in appropriate finite dimensional
($\Z_2$-graded) bundles. The physical interpretation of the
construction is through the appearance of $D$-branes of higher
codimension in the process of annihilation of
$D-\overline{D}$-branes with the corresponding nontrivial
low-dimensional RR charges. In order to get an arbitrary brane
configuration one could use only $D$-branes of the highest
possible dimension. For simplicity we will illustrate these
phenomena looking at type IIB theory and $D9-\overline{D9}$ system
\cite{WitK}. It should be noted that the configurations with non
equal number of $D9$ and $\overline{D}9$ branes is anomalous in
string theory. But in our approach to RR-couplings all $D$ branes
are more or less on  equal footing and we will not discriminate
the case of $D9$ branes in the following presentation.

As an
application we derive explicit formula for the topological density
of the "fat" $D5$-brane inside the  $D9$-brane, which coincides
with the well known formula for the densities of the instanton
charge (see e.g. \cite{CorGod}). The main deficiency of the
approach of this section is the necessity to consider special
configurations of the $D9-\overline{D9}$-branes depending on the
desired ultimate configuration of the lower dimensional
brane/anti-brane system we would like to get. The lesson learned
from these examples suggests the idea to search for the universal
description by taking the number of $D9-\overline{D9}$ branes
infinite \cite{WitN} (see \cite{Hori} for earlier discussions)
\footnote{Another motivation for this comes from the search of
closed string in the field theory of open strings \cite{Sh}- $N
D1$-branes thought as multi-soliton on space-filling $D9$ brane,
or as $D9$ compactified on $T^8$, in large $N$ limit and IR
describes type II closed string via Matrix String construction
\cite{DVV}.}.

Our main proposal is presented in Section 3. Here we
describe the concrete procedure of taking the limit by considering
the superconnection in the appropriate geometrically defined
infinite dimensional bundle. We show how the explicit formulas
from the section 2 appear in this formalism. At the end we briefly
discuss the interpretations of the proposed universal description
in terms of string theory. The issues related to the appearance of
non-abelian Chan-Paton factors in the proposed universal
description and its connection with background independent open
string field theory \cite{WBI}, \cite{SBI} will be published
separately \cite{ First}.

\section{RR-couplings from the finite number of
$D9-\overline{D9}$-branes}

In this section we consider the RR gauge field couplings with open
strings for the backgrounds with the finite number
$D9-\overline{D9}$-branes and the low dimensional brane
backgrounds arising in the process of the annihilation.

The
general coupling of the $D$-brane fields with the RR fields may be
characterized as follows. Any configuration of $D$-branes in type
IIB theory defines some element of the $K^0$-group of space-time
\cite{Moore,WitK} which may be represented as a formal difference
of the vector bundles up to some equivalence. This was interpreted
in \cite{WitK} as gauge bundles on $D9$ and $\overline{D9}$ branes
filling whole space-time. There is a canonical map of the
$K^0$-group to the cohomology of the manifold defined with the
help of the Chern character. In order to get explicit
representation of this map in terms of the (closed) differential
forms one should supply the difference of the bundles with an
appropriate analog of the connection --- superconnection
\cite{Quillen}. The superconnection corresponding to a given
configuration of $D9-\overline{D9}$-branes may be constructed in
terms of the open string field. The coupling of the RR gauge
fields with open string sector of the $D9-\overline{D9}$-branes
filling whole space-time has the following form: \bqa
\label{coupling} S_{RR}= \mu_9\int_{M^{(10)}} [C_{RR} \wedge {\rm
Ch}(\A)]_{top} \, \eqa Here $\mu_9$ is the $D9$- and
$\overline{D9}$-brane charge, $M^{(10)}$ is ten-dimensional flat
target space, $Ch(\A)$ is the Chern character of the
superconnection $\A$ constructed from the open string modes,
$C_{RR}$ is defined as a sum of the RR gauge fields of the given
parity (even in IIB theory): \bqa\label{modes} C_{RR}=\sum_{k}
C_{(2k)}, \quad C_{(2k)} = C_{\mu_1\dots\mu_{2k}} \, dx^{\mu_1}
\wedge \dots \wedge dx^{\mu_{2k}} \eqa and the subscript $top$
means that we should integrate the top-dimensional differential
form over $M^{(10)}$. We use the normalization of RR gauge fields
consistent with the standard definition of the superconnection
used below.

The gauge bundles on $D9$ and $\overline{D9}$ branes
may be combined in a $\Z_2$ graded bundle with the superconnection
defined in terms of the gauge field $A_{\mu}$ on the $D$, the
gauge field $\widetilde{A}_{\mu}$ on the $\overline{D}$ and the
complex tachyon field $T$ corresponding to the lowest energy
excitation of the strings stretched between $D$ and $\overline{D}$
by the following:

\bqa \label{superconnection} \A = \left(
\begin{array} {c c}
dx^{\mu} \, \nabla^+_{\mu} & T \\ \overline{T} & dx^{\mu} \,
\nabla^-_{\mu}
\end{array}
\right), \nonumber \\ \nabla^+_{\mu} = \partial_{\mu} + {\rm i} \,
A_{\mu}, \quad {\rm and} \quad \nabla^-_{\mu} = \partial_{\mu} +
{\rm i} \, \widetilde{A}_{\mu}.
\eqa
The Chern character of the
superconnection is defined by the standard formula:

\bqa {\rm
Ch}(\A) = {\rm Str} \, e^{- \frac{\A^2}{2\pi}}.
\eqa
Here we use
the supertrace on the matrices acting on $\Z_2$-vector spaces
which is defined as follows. Let $V=V_+\oplus V_-$ be a
$\Z_2$-vector space and $\mathcal{O}\in End(V)$:

\bqa
\begin{pmatrix}
\mathcal{O}_{++} & \mathcal{O}_{+-} \\ \mathcal{O}_{-+} &
\mathcal{O}_{--}
\end{pmatrix}. \eqa
Then the  supertrace of $\mathcal{O}$ is defined as: \bqa {\rm
Str}_{V} \, \mathcal{O} = {\rm Tr}_V \, \tau \, \mathcal{O} = {\rm
tr}_{V_+} \, \mathcal{O}_{++} - {\rm tr}_{V_-} \mathcal{O}_{--},
\eqa where $\tau$ is an operator defining $\Z_2$-structure.

In the
following subsections we consider various families of the
background open string fields and show how the RR gauge field
couplings for the lower dimensional $D$-branes appear in some
limits.

\subsection{$D$-instantons from $D9-\overline{D9}$-branes}
We start with the simple case of $k$ $D$-instantons inside
ten-dimensional flat space, which appear as the result of the
$D9-\overline{D9}$ annihilation. In order to construct $k$
$D$-instantons from the $D9-\overline{D9}$ system one has to
consider $16 \, k$ $D9$-branes and $16 \, k$
$\overline{D9}$-branes \cite{Sen,WitK,Hor}, where 16 Chan-Paton
indexes from both sides are embedded into the spinor bundles of
the target space with opposite chirality.

The corresponding
superconnection may be described as follows \cite{WitK}. Consider
a $\Z_2$-graded vector bundle $V=V_+ \oplus V_-$, where
$V_{\pm}=S_{\pm}\otimes E_k$ and $S_{\pm}$ are the spinor bundles
of the definite chirality and $E_k$ is a $k$ -dimensional vector
bundle. The superconnection is defined as in
(\ref{superconnection}) with:

\bqa \nabla^+ : V_{+} \rightarrow V_{+} \nonumber \\ \nabla^- :
V_{-} \rightarrow V_{-} \nonumber \\ T : V_{+} \rightarrow V_{-}
\nonumber \\ \overline{T} : V_{-} \rightarrow V_{+} \eqa We take a
trivialization of the vector bundle $E_k$, i.e. $\nabla^{\pm}= d$.

To begin we describe the simplest situation of coincident
$D$-instantons which are left after the $D9-\overline{D9}$
annihilation. The corresponding superconnection may be constructed
as the limit $t \rightarrow 0$ of the superconnection
(\ref{superconnection}) with: \bqa \label{tachsol} \nabla = d
\quad {\rm and} \quad T = \frac{1}{\sqrt{t}} \, x^{\mu} \,
\sigma_{\mu} \otimes \mathbf{1}_{k\times k}, \eqa where
$\sigma_\mu$ are ten-dimensional $\gamma$-matrices in the
Majorana-Weyl representation and $\mathbf{1}_{k\times k}$ is the
identity matrix acting in the space $E_k$ ($k$ is the number of
the instantons). Description of Ch$(\A)$ in terms of such a family
of superconnections is a standard procedure and appeared in
mathematical literature long ago \cite{Quillen}. In the language
of  string theory it may be interpreted as off-shell interpolation
(one shall compare this tachyon profile $T$ with the one from
\cite{WitK}). The family (\ref{tachsol}) corresponds to the
smearing of the $D$-instanton charge over a region of the size
$\propto \sqrt{t}$ (see eq. (\ref{delta}) below).

The square of
the superconnection in question has the form:

\bqa \A_t^2 = \left(\begin{array} {c c} \frac{1}{t} \, |x|^2 &
\frac{1}{\sqrt{t}} \, dx^{\mu} \, \sigma_{\mu} \\
\frac{1}{\sqrt{t}} \, dx^{\mu} \, \overline{\sigma}_{\mu} &
\frac{1}{t} \, |x|^2
\end{array} \right) \otimes \mathbf{1}_{[k\times k]} =
\left(\frac{1}{\sqrt{t}} \, dx^{\mu} \, \gamma_{\mu} + \frac{1}{t}
\, |x|^2 \right) \otimes \mathbf{1}_{[k\times k]}. \eqa Taking
into account that in the spinor notations supertrace of the
arbitrary operator $\mathcal{O}$ may be represented in terms of
the trace over spinor representation, $Sp$, as \bqa {\rm Str}
\mathcal{O} = {\rm tr_k}\, {\rm Sp} \, \gamma^{11} \mathcal{O},
\eqa we have: \bqa {\rm Ch}(\A_t) = k \, {\rm Sp} \, \gamma^{11}
e^{ - \frac{\A_t^2}{2\pi}}. \eqa Here $k=\,$tr$\,
\mathbf{1}_{[k\times k]}$. Thus, using the identities for d
dimensional space: \bqa {\rm Sp} \gamma^{d+1} \, \gamma^{\mu_1}
\dots \gamma^{\mu_d} = (2i)^{\frac{d}{2}}\epsilon^{\mu_1 \dots
\mu_d} \eqa and \bqa \lim_{t\rightarrow 0} \, \frac{1}{(\pi
t)^{d/2}} \, \exp\left\{-\frac{|y|^2}{t}\right\} =
\delta^{(d)}(y), \label{delta} \eqa we obtain: \bqa {\rm Ch}(\A) =
\lim_{t\rightarrow 0} \, {\rm Ch}(\A_t) = k \, \delta^{(10)}(x) \,
vol(M^{10}). \eqa After the substitution into (\ref{coupling})
this gives $S_{RR} = k \, \mu_{-1} \, C_{(0)}(x=0)$, i.e. the
correct coupling of $k$ $D$-instantons with RR gauge fields.

Consider a more general situation given by the deformation of the
tachyon profile (\ref{tachsol}) as follows: \bqa \label{tprofile}
T = \frac{1}{\sqrt{t}} \, \left(x^{\mu\phantom{\frac12}}
\gamma_{\mu} \otimes \mathbf{1}_{k\times k} - \Phi^{\mu} \,\,
\gamma_{\mu}\right), \eqa with $\Phi^{\mu}$ - constant $k\times k$
hermitian matrix, $\mu = 0, ..., 9$, which capture the
fluctuations of the relative positions of $k$ $D$-instantons.
These matrices naturally appear in the low energy descriptions of
the $D$-instantons in string theory \cite{WitDbr}. For the square
of the superconnection we have: \bqa \label{dinst}
\A_t^2=\frac{1}{\sqrt{t}} \, dx^{\mu} \, \gamma_{\mu} \otimes
\mathbf{1}_{k\times k} + \frac{1}{2t} \, [\Phi^{\mu}, \,
\Phi^{\nu}] \gamma_{\mu}\, \gamma_{\nu} + \frac{1}{t} \, |x^{\mu}
\otimes \, \mathbf{1}_{k \times k} \,\, - \,\, \Phi^{\mu}|^2. \eqa
Let us define the symmetric trace "SymTr" following
\cite{Tseytlin,Myers} as the symmetrisation of the usual trace.
Thus we have: \bqa \label{simtr} {\rm Tr} \, e^{A + B + C} = {\rm
Tr} \, \sum_n \frac{\left(A + B + C\right)^n}{n!} = {\rm SymTr} \,
\sum_n \frac{\left(A + B + C\right)^n}{n!} = {\rm SymTr} \, e^{A +
B + C} \eqa Taking into account (\ref{dinst}) and (\ref{simtr}) we
obtain: \bqa \label{myers1} S_{RR} &=& \mu_{-1} \, \int_{M^{(10)}}
C_{RR}(x) \, {\rm SymTr} \, e^{- [i_{\Phi}, \phantom{\frac12}
i_{\Phi}]/2\pi} \, \delta^{(10)}(x \otimes
\mathbf{1}^{\mu\phantom{\frac12}} - \,\, \Phi^{\mu}) = \nonumber
\\ &=& \mu_{-1} \, \int_{M^{(10)}} C_{RR}(x) \, {\rm SymTr} \, e^{ -
[i_{\Phi}, \phantom{\frac12} i_{\Phi}]/2\pi} \, e^{-{\rm
i}\Phi_{\mu} \pr_{\mu}} \,\delta^{(10)}(x) = \nonumber
\\ &=& \mu_{-1} \, {\rm SymTr} \, e^{ -
[i_{\Phi}, \phantom{\frac12} i_{\Phi}]/2\pi}\, C_{RR}(\Phi) \eqa
where $i_{\Phi}$ is defined in \cite{Myers} - the indexes of
$\Phi$'s in the commutator in the exponent are contracted with
those of $C_{(2k)}$. We also use the following definition of the
$\delta$-function: \bqa \delta^{(d)}(x^{\mu} \otimes
\mathbf{1}^{\phantom{\frac12}} - \,\, \Phi^{\mu}) = \lim_{t\to 0}
\, \frac{1}{{(\pi t)}^{d/2}} \, \exp \{- \frac{1}{t} \,
\sum^{d}_{\mu=1} |x^{\mu} \otimes \mathbf{1}^{\phantom{\frac12}} -
\,\, \Phi^{\mu}|^2\}. \eqa Note that the expression (\ref{myers1})
coincides with $S_{RR}$ obtained in \cite{Myers}.

One could
consider more general deformations of $T$: \bqa \label{TTT} T
\propto x^{\mu} \, \sigma_{\mu} \otimes \mathbf{1} +
T_{(k)}^{\mu_1 \cdots \mu_k}(x) \, \sigma_{\mu_1 \cdots \mu_k} +
\cdots, \eqa which apparently correspond to massive open string
excitations.

\subsection{$Dp$-branes from the $D9-\overline{D9}$}
It is not hard to generalize the above construction to the case of
$k$ $Dp$-branes with $p < 9$. Let us separate indexes $\mu =
0,...,9$ into $m = 0,...,p$ and $i = p+1,...,9$ and consider the
following background values of the tachyon and gauge field: \bqa
\label{tachsol1} \nabla = d \quad {\rm and} \quad T =
\frac{1}{\sqrt{t}} \, x^{i} \, \sigma_{i} \otimes
\mathbf{1}_{k\times k}. \eqa Here now $\sigma_i$ are
($9-p$)-dimensional $\sigma$-matrices. Now, in order to construct
$k$ $Dp$-branes from the $D9-\overline{D9}$ system, one has to
consider \cite{WitK} $2^{(9-p)/2}\times k$ $D9$-branes and the
same number of $\overline{D9}$-branes.

Similarly to the case of
the instantons the superconnection (\ref{tachsol1}) leads to the
coupling: $$ S_{RR} = k \mu_{p} \int_{M^{(p+1)}} C_{(p+1)} (x_m),
$$ which is the source corresponding to $k$ $Dp$-branes;
$M^{(p+1)}$ here is the world-volume of the $Dp$-branes.

Let us
consider the fluctuations around (\ref{tachsol1}) of the form:
\bqa \nabla = \left[\pr_m \, + \, {\rm i}^{\phantom{\frac12}}
A_m(x_m)\right] \, dx^m + \pr_i \, d x^i \quad {\rm and} \quad T =
\frac{1}{\sqrt{t}}\, \left[x^{i\phantom{\frac12}} \sigma_{i}
\otimes \mathbf{1}_{k\times k} - \Phi^{i}(x_m) \,\, \sigma_{i}
\right]. \label{tachval} \eqa Here $A$ and $\Phi$ are the gauge
field and scalar field living on the $Dp$-brane world
volume\footnote{Note that while the scalars $\Phi_i$ appear from
the tachyon field, the gauge field $A_m$ in this case originates
from the gauge field localized on the $D9-\overline{D9}$ system.
In the next section we will consider the "gauge equivalent"
description where $A_m$ will take its origin from the tachyon
field as well.}. The explicit expression for the square of the
superconnection is: \bqa \A^2_t &=& [\nabla_m, \phantom{\frac12}
\nabla_n ] \, dx^m \, dx^n + \frac{1}{\sqrt{t}} \, dx^{i} \,
\gamma_{i} \otimes \mathbf{1}_{k\times k} + \frac{1}{\sqrt{t}}\,
[\nabla_m, \phantom{\frac12} \Phi_i ]\gamma^i \, dx^m + \nonumber
\\ &+& \frac{1}{2t} \, [\Phi^{i}, \phantom{\frac12} \Phi^{j}]
\gamma_{i}\, \gamma_{j} + \frac{1}{t} \, |x^{i}\otimes
\mathbf{1}_{k\times k}^{\phantom{\frac12}} -\,\, \Phi^{i}|^2. \eqa
After substitution of this expression into (\ref{coupling}) and
taking $t \to 0$, one finds: \bqa \label{myers2} S_{RR} = \mu_p \,
\int_{M^{(p+1)}} \left[{\rm SymTr} \, C_{RR}(x_m, \, \Phi_i) \,
e^{ - \frac{1}{2\pi}\left(F_{(2)}(x_m) + [\nabla_{(1)}(x_m),
\phantom{\frac12} i_{\Phi} ] + [i_{\Phi}, \phantom{\frac12}
i_{\Phi}]\right)}\right]_{top}, \eqa with $F_{(2)} = [\nabla_m, \,
\nabla_n] \, dx^m \wedge dx^n$ and $\nabla_{(1)} = \nabla_m \,
dx^m$. This expression also coincides with $S_{RR}$ considered in
\cite{Myers}.
\subsection{General configuration of $D$-branes of various codimensions}
The general case is now rather obvious. For illustration we will
construct a very simple configuration of low dimensional
$D$-branes. Consider $D$-branes oriented along coordinate
hyperplanes in $d=10$ dimensional flat space. One describes such a
configuration with $k_p$ $Dp$-branes and $\widetilde{k}_p$
$\overline{Dp}$-branes for all $p$  by considering the collection
of $D9-\overline{D9}$ which represents $\Z_2$-graded bundle:
\bqa\label{general1} V=V_+\oplus V_-=\oplus_{p} (E_{k_p}\oplus
E_{\widetilde{k}_p})\otimes S(9-p). \eqa Here $E_{k_p}$ and
$E_{\widetilde{k}_p}$ are vector bundles of dimension $k_p$ and
$\widetilde{k}_p$ and $S(9-p)$ are the spinor bundles in the
directions transverse to the $Dp$-branes ($S(0)$ is a trivial
$\Z_2$-even bundle). The sum in (\ref{general1}) is taken over the
odd $p$'s (we are considering type IIB string theory).
$\Z_2$-grading is given by the operator $\tau$ acting on spinor
bundles $S(9-p)$ as $\tau = (-1)^{p/4} \, \gamma^{10-p}$ where
$\gamma^{10-p}$ is a product of all gamma-matrices acting in
$S(9-p)$ (analog of $\gamma^5$ in $(9-p)$ dimensions). It acts on
the bundles $E_{k_p}$ and $E_{\widetilde{k}_p}$  as follows: \bqa
\tau E_{k_p} &=& E_{k_p} \nonumber \\ \tau E_{\widetilde{k}_p} &=&
-E_{\widetilde{k}_p} \eqa

Then the obvious generalization of the superconnection
(\ref{tachval}) is: \bqa \A_{t,\widetilde{t} } = \nabla &+& \sum_p
\frac{1}{\sqrt{t_p}} \sum_{i=p+2}^{10} (x^{i\phantom{\frac12}}
\gamma_{i} \otimes {\bf 1}_{k_p\times k_p} - \Phi^i \, \gamma_i
)+\\ &+& \sum_p \frac{1}{\sqrt{\widetilde{t}_p}} \sum_{i=p+2}^{10}
(x^{i\phantom{\frac12}} \gamma_{i} \otimes {\bf
1}_{\widetilde{k}_p\times \widetilde{k}_p} - \widetilde{\Phi}^i \,
\gamma_i). \label{nablanabla} \eqa Here the matrices ${\bf
1}_{k_p\times k_p}$ (${\bf 1}_{\widetilde{k}_p\times
\widetilde{k}_p}$) acts as unit matrices on $E_{k_p}$
($E_{\widetilde{k}_p}$) and by zero otherwise. It gives rise to
the density of the Chern character which in the limit $t_p
,\widetilde{t}_p \rightarrow 0$ describes the coupling of the
collection of the $Dp$-branes with the RR fields.

The general superconnection (\ref{nablanabla}) interpolates
between various configurations of D-branes capturing all their
low-energy fluctuations. Note an interesting novel feature of the
general case. The general deformations of this superconnection
includes operators transforming in the {\em spinor}
representations of the transversal Lorentz groups. These tachyonic
excitations correspond to the lowest energy modes of the strings
stretched between the $Dp$- and $\overline{Dp}$-branes with the
mixed Neumann-Dirichlet boundary conditions. They are responsible
for the "thickening" of the lower dimensional $D$-branes inside
bigger $D$-branes. In the next subsection we proceed with the
discussion of some example of this situation.

\subsection{ $D5$-branes inside $D9$ -branes and the instanton charge
density}

Consider the special case of the $k$ $D5$-branes inside
$N$ $D9$-branes as the result of the $D9-\overline{D9}$
annihilation. The theory with $N>0$ is anomalous. However, this is
not important for our further considerations and we proceed with
this case in order to make the presentation most transparent.
Otherwise one could consider $D3-D7$-brane system using the
machinery of the previous subsection.

The $\Z_2$-graded vector
bundle in this case has the form: \bqa V=V_- \oplus V_+ =
(W^{(9)\phantom{\frac12}}_N \oplus \,\, E^{(5)}_k \otimes
S(4)_-)\, \, \oplus \, \, (E^{(5)}_k \otimes S(4)_+), \eqa where
$S_{\pm}$ are definite chiral, Weyl fermion bundles in four
dimensional space transverse to the $D5$-branes. We take a
trivialization of the $E^{(5)}_k$ bundles $\nabla = d$.

In order
to make a comparison with on-shell description in terms of the
instanton moduli space we restrict the superconnection under
consideration by the condition that it is invariant with respect
to the natural action of $Sp(1)$-group on $S_+$. Then its moduli
space is parameterized by ADHM data (see e.g. \cite{DonK}): \bqa T
& = & \frac{1}{\sqrt{t}} \Delta \nonumber
\\ \Delta & : & S_-(4)\otimes E_k^{(5)} \oplus W_N^{(9)}
\rightarrow S_+(4) \otimes E_k^{(5)}. \eqa Here $\Delta$ is the
standard ADHM matrix \cite{DonK}: \bqa \Delta =
[x_i^{\phantom{\frac12}} \sigma^i \otimes \mathbf{1}_{k\times k} +
B_i(x_m)\, \sigma^i, \, h(x_m)], \eqa where $i=6,...,9$ are
directions transversal to the $D5$-brane and $x_i \, \sigma^i$
naturally acts as the operator $x : S_-\rightarrow S_+$, similarly
constant (independent of $x_i$) operator $B$ acts as $B :
S_-\otimes E_k \rightarrow S_+\otimes E_k$ and the constant
operator $h$ acts as $h : W_N\rightarrow S_+\otimes E_k$. In the
language of string theory $B_i$'s are scalars localized on the
$D5$-brane world-volume, which appear as the massless excitations
of the strings with both their ends on the $D5$-brane. They
parameterize transversal fluctuations of the $D5$-brane. In these
 terms $h$ appears as the massless
excitations of strings stretched between $D5$- and $D9$-branes.

The condition on the matrix $\Delta$ in ADHM construction is
equivalent to invariance with respect to the action of
$Sp(1)$-group on $S_+$ and thus we cover the whole ADHM moduli
space. The derivation of this tachyon soliton from the condition
of SUSY restoration after the annihilation was given in
\cite{Akhmedov}.

The Chern character of the superconnection $\A_t$ at the limit
$t\rightarrow 0$ reduces to the standard expression for the Chern
character of the instanton solutions. Consider for the
illustration the simple case of one instanton in the $Sp(1)$ gauge
theory: $k=1$, dim $W=2$. It is useful to identify $W\cong S_+$.
Choose the configuration of the instanton with the center at $x=0$
and of the radius $h$. Then the corresponding superconnection is:

\bqa \A_t = d \otimes \mathbf{1}_{4\times 4} + \frac{1}{\sqrt{t}}
\,\left[x^i \, \gamma_i \oplus 0_W + h \,\, p_w +
h^{\phantom{\frac12}} p_w^* \right]. \eqa Here $B=0$ and $p_w:
W\rightarrow S_+$ identifies $W$ and $S_+$. Note that
$\frac{1-\gamma^5}{2}p_w=p_w$ and $\frac{1+\gamma^5}{2}p_w=0$. The
square of the superconnection is: \bqa \A_t^2 = \frac{1}{\sqrt{t}}
\, dx^{i} \, \gamma_i \oplus 0_W + \frac{1}{t} \,
\left[|x|^{2^{\phantom{\frac12}}} \mathbf{1}_S + h^2 \,
\mathbf{1}_{S_+} + h^2 \, \mathbf{1}_W + h \, x^{i} \, \sigma_i \,
p_W + h \, p_W^* \, x^{i} \, \overline{\sigma}_i \right], \eqa
where we use the notations: $\mathbf{1}_S$ is the projector on the
spinor subspace, $\mathbf{1}_{S_+} = \frac{1 + \gamma^5}{2}$ is
the projector on the $S_+$ subspace and $\mathbf{1}_W$ is the
projector on $W$. Thus, in matrix form, such {\bf A}
superconnection looks as: \bqa \A_t^2 = \left(\begin{array} {c c
c} \frac{1}{t} \,\left(|x|^2 + h^2\right) & \frac{1}{\sqrt{t}} \,
dx^{i} \, \sigma_{i} & 0 \\ \frac{1}{\sqrt{t}} \, dx^{i} \,
\overline{\sigma}_{i} & \frac{1}{t} \, |x|^2 & \frac{1}{t} \, h \,
x^i \sigma_i \\ 0 & \frac{1}{t} \, x^i \, \overline{\sigma}_i \, h
& \frac{1}{t} \, h^2
\end{array} \right).
\eqa One can use the identity\footnote{In order to prove this
formula consider "log"-derivative over $t$ of the operator
$G(t)=e^{t \, (A + B)}e^{- t \, B}$: \bqa G(t)^{-1}d_t G(t) =
e^{-t \, B} \, A \, e^{t \, B} \nonumber \eqa The integration over
$t$ gives the desired identity.}: \bqa \label{exprrr} e^{A + B} =
{\rm P}\, \exp\left\{\int^1_0 dt \, e^{-t\, B} \, A e^{t\,
B}\right\} \, e^B \eqa and obvious commutation relations between
$\gamma_i$, $\mathbf{1}_S$, $\mathbf{1}_{S_+}$, $\mathbf{1}_W$ and
$p_w$, in order to expand the expression $\exp (-\A_t^2)$ in the
components having the differential forms of the definite degree
and prove that:

\bqa {\rm Ch}(\A) = \lim_{t\rightarrow 0} \, {\rm Sp} \, {\rm tr}
\, (\gamma^5 \oplus \mathbf{1}_W )\, e^{-\frac{\A_t^2}{2\pi}}
\propto {\rm dim}(W) +\frac{1}{8\pi} \frac{h^4}{(|x|^2 + h^2)^4}
\, vol(\R^4). \eqa This is the correct expression for the
topological charge density of one $Sp(1)$ YM instanton
\cite{CorGod}. The more conceptual derivation of this formula and
its generalizations to the multiple instantons with higher range
gauge groups using the results of \cite{GerKot} will be published
elsewhere \cite{GerKot1}.

\section{Universal description and algebra of differential operators}

The construction described in the previous section has an obvious
drawback. Auxiliary (super)-vector bundles that appear in the
description of $D$-branes do not have clear geometrical meaning.
In particular for each $D$-brane configuration one needs to
consider the specific collection of vector bundles. On the other
hand it is obvious that different vector bundles with the same
topological invariants lead to the equivalent description of the
$D$-branes. In order to get a universal description of $D$-brane
backgrounds it is natural to look at the bundles of the infinite
rank \cite{WitN}. We will use a concrete realization of this
limiting bundle as the bundle where fiber is identified with the
space of functions (more exactly sections of the
finite-dimensional bundles) on an auxiliary space. As an auxiliary
space it appears natural to take a copy of the space where string
theory is defined. The idea to use this realization of the
universal bundle comes from simple qualitative arguments
concerning the configuration space of the open strings that were
discussed in the introduction. Note that the natural algebra
acting on the fiber of this bundle is the algebra of differential
operators on auxiliary space.

This construction may be considered
as an application of the general approach of \cite{First} to the
description of the RR gauge field couplings.

In the next
subsection we show how to derive the corresponding RR field
coupling for the $D9$-brane filling the whole space-time. Then we
proceed with the derivation of the RR gauge field coupling for an
arbitrary case. This reproduces the expressions considered in
Section 2.

\subsection{Description of $D9$ branes}

We start the description of the universal construction with simple
physical motivations. Consider $16\, k$ $D9$-branes and $16\, k$
$\overline{D9}$-branes in the flat space -time $M=\R^{10}$ and
take $k\to\infty$. In this way we construct $k\to\infty$
$D$($\overline{D}$)-instantons. We could use these
$D$($\overline{D}$)-instantons to construct various $Dp$-branes.
The configuration of $k$ non-coincident $D$-instantons may be
parameterized by $k$ points of space-time. More correctly it is
the space of $10$ mutually commuting $k \times k-$matrices up to
conjugation. It is easy to see that as the number $k$ of
$D$-instantons tends to infinity the $k$-dimensional vector space
on which these matrices act becomes more and more like Hilbert
space of the square integrable functions on space-time $M$ where
string theory is defined. In these terms the matrices themselves
may be considered as differential operators (more generally
integral operators) acting on functions on $M$. The traces of
matrices in this limit may be reduced to the integrals over the
space $M$. For any operator function $F(\widehat{q},\widehat{p})$
of the coordinates $\widehat{q}^{\mu}$ and momentum
$\widehat{p}_{\mu}$ one has: \bqa {\rm Tr}_{\mathcal{H}}
:F(\widehat{q},\widehat{p}):= \int dp \wedge dq \left<
q|:F(\widehat{q},\widehat{p}):|p \right> \left< p|q \right> = \int
dp \wedge dq \, F(q,p), \eqa where $:F:$ is the normal ordering of
the operators such that the momentum operators act on the right
and the position operators act on the left.

Note that to have a
well defined trace the operators should have special property (to
be of the trace class). A particular class of such operators is
given by the expressions $F(\widehat{p},\widehat{q})
\propto\delta^{(10)}(\widehat{p})f(\widehat{q})$. It's trace is
naturally represented by the integral over the lagrangian
submanifold $M\in T^*M$ rather than over $T^*M$ and thus the
operators of this  kind and its smooth deformations are natural
candidates for the limits of the appropriate matrix variables.
This type of operator has a simple qualitative interpretation in
terms of string backgrounds. Let us symbolically denote open
string by the  matrix $K_{x,y}$ (or more exactly integral
operators $\widehat{K}$ with the kernel $K(x,y)$ acting on the
space of functions on $M$) where possible end points of the
strings are enumerated by the indexes "$x$" and "$y$". The trace
invariants naturally correspond to the open strings with
identified ends and thus may be considered as closed strings. Now
let us consider the trace of the operator $\widehat{K}$ with
additional insertion of the operator $\delta(\widehat{p})$
($\widehat{p}$ is momentum operator) which projects on the states
with zero eigenvalue of $\widehat{p}$. It is easy to see that:
\bqa \delta(\widehat{p})=|p=0\left>\right< p=0|=(\sum_x |x \left
> \right. )(\sum_{x'} \left < x'|)\right.  \eqa (note that the identity operator is
$1=\sum_x |x \left>\right<x|$) and we have: \bqa {\rm
Tr}_{\mathcal{H}} \,
 \delta(\widehat{p})\widehat{K}\sim \sum_{x,x'} K(x,x') \eqa We see
that in the presence of the operator we should sum over the
positions of the "ends" of the string independently. Thus the
insertion of the operator $\delta(\widehat{p})$ may be interpreted
as the creation of the $D$-brane filling the whole space-time.

We have replaced the consideration of the infinite number of
$D$-branes by the consideration of the infinite-dimensional vector
bundles over the space-time $M$ with the fiber --- the space of
functions on the copy $\widetilde{M}$ of $M$. The Lie algebra of
the differential operators acts naturally in the fibers of the
bundle. Note that one may consider infinite dimensional bundles
with the fiber given by the sections of a finite dimensional
(super)-bundle. We encounter examples with the space of sections
of spin bundles as a fiber.

Now we give an explicit construction of the superconnection in
this bundle which leads to the desired description of
RR-couplings. Let us start with few remarks on the notations. In
order to distinguish the space-time $M$ from the auxiliary space
$\widetilde{M}$ we will use the coordinates $(x^{\mu})$ on
$M\equiv M_x$ and the coordinates $(y^{\mu})$ on
$\widetilde{M}\equiv M_y$. Thus the matrices realize the
quantization of the space of functions on $T^*M_y$ and the
corresponding Hilbert space may be constructed in terms of the
sections of the spinor bundle over $M_y$.

Having in mind more general cases treated below we consider
Clifford algebra for the total space $M_x\otimes T^*M_y$: \bqa
\label{Clif1} \{\gamma_{\mu},\gamma_{\nu}\}=\delta_{\mu \nu} \quad
\{\Gamma_{\mu},\Gamma_{\nu}\}=\delta_{\mu \nu} \quad
\{\widehat{\Gamma}^{\mu},\widehat{\Gamma}^{\nu}\}=\delta^{\mu \nu}
\nonumber \\
\{{\Gamma}_{\mu},\gamma_{\nu}\}=\{\widehat{\Gamma}^{\mu},\Gamma_{\nu}\}=
\{\widehat{\Gamma}^{\mu},\gamma_{\nu}\}=0. \eqa The gamma matrices
are defined with respect to the explicit coordinates as follows:
$(x^{\mu},y^{\nu},p_{\rho}) \leftrightarrow
(\gamma_{\mu},\Gamma_{\nu},\widehat{\Gamma}^{\rho})$

First we give a simple example of the superconnection describing
the  $D9$-brane filling the whole space-time. Taking into account
the heuristic description in terms of the infinite number
$D$-instantons considered at the beginning of this subsection and
the results of the first part of the paper we could propose the
following superconnection: \bqa\label{tremss} \A_{s,t} = d +
\frac{1}{\sqrt{t}} \, \gamma_{\mu}(x^{\mu}-y^{\mu}) +
\frac{1}{\sqrt{s}} \, \widehat{\Gamma}^{\mu} \,
\frac{\partial}{\partial y^{\mu}} \eqa This superconnection takes
values in the space of differential operators acting in the
auxiliary space of sections of spinor bundle on $M_y$ with
standard $\Z_2$-structure defined by the chirality. The first term
is a direct analog of the tachyonic profile (\ref{tprofile}). Note
also that the last term is a Dirac operator on this auxiliary
space.

The superconnection (\ref{tremss}) corresponds to the
tachyon field of the following form: \bqa T\left(x
|^{\phantom{\frac12}} y,\widehat{p}\right) = \frac{1}{\sqrt{t}} \,
\sigma_{\mu}(x^{\mu}-y^{\mu}) + \frac{1}{\sqrt{s}}\,
\widehat{\Sigma}^{\mu} \widehat{p}_{\mu} \eqa where $\widehat{p} =
\pr/\pr y$. The first term is rather obvious and one of the
reasons for introducing the last term is to fulfill the condition
of being trace class.

The square of (\ref{tremss}) is given by the
expression: \bqa \A_{s,t}^2=\frac{1}{t}\, |x-y|^2+\frac{1}{s}\,
\left|\widehat{p}_{\mu}\right|^2 +\frac{1}{\sqrt{st}} \,
\gamma_\mu \, \widehat{\Gamma}^\mu + \frac{1}{\sqrt{t}} \,
dx^{\mu} \, \gamma_\mu \eqa and the  Chern character is given by:
\bqa {\rm Ch}(\A) = \lim_{t\rightarrow 0} \lim_{s\rightarrow
0}{\rm Ch}(\A_{s,t}) \eqa Note that here "Str" is taken over the
representation of the full Clifford algebra (\ref{Clif1}) and
includes ${\rm Tr}_{\mathcal{H}}$ as well. By the trivial
considerations it may be reduced to the following trace over the
space of $y$-dependent functions: \bqa {\rm Ch}(\A) = {\rm
Tr}_{\mathcal{H}} \, \delta^{(10)}\left(\widehat{p}^{\mu}\right)
\, \delta^{(10)} (x-y) = 1 \eqa Thus we have reproduced the
correct coupling with the top-degree RR-form for the $D9$-brane
filling whole space-time: $S_{RR} = \mu_9 \, \int_{M^{(10)}} C_{(10)}(x)$.

One could include a non-trivial connection over $y$-variables. Let
us show that this is equivalent to the inclusion of the same
connection on $M_x$. Consider the following superconnection:

\bqa
\A_{s,t}=d + \frac{1}{\sqrt{t}} \, \gamma_{\mu}\, \left(x^{\mu} -
y^{\mu}\right) + \frac{1}{\sqrt{s}} \, \widehat{\Gamma}^{\mu} \,
\left(\partial_{\mu} + {\rm i}^{\phantom{\frac12}} A_\mu(y)\right)
\eqa
Simple calculation gives:

\bqa \label{aaa} {\rm Ch}(\A) =
{\rm Tr}_{\mathcal{H}} \, \left[\delta^{(10)}(\pr_{\mu} \, + \, {\rm
i}^{\phantom{\frac12}} A_{\mu}) \, \delta^{(10)}(x-y) \, \, \,
\exp\left\{- \frac{1}{2\pi}F_{(2)}[A(y)]\right\} \right] = \nonumber
\\ = \exp\left\{-\frac{1}{2\pi} F_{(2)}[A
(x)]\right\}
\eqa
At the last step we use the following identity: let us
represent $\delta$-function as:

\bqa \delta^{(10)} \left(\pr_{\mu}
\,+ \, {\rm i}^{\phantom{\frac12}} A_{\mu}(y)\right) = \int dq \,
e^{{\rm i} \, q^\mu \, \left(\pr_{\mu} \,\, + \,\, {\rm
i}^{\phantom{\frac12}} A_{\mu}(y)\right)} \eqa and use: \bqa
\label{ident2} e^{{\rm i} \, q^\mu \, \left(\pr_{\mu} \,\, + \,
\,{\rm i}^{\phantom{\frac12}} A_{\mu}(y)\right)} = e^{{\rm i} \,
q^{\mu} \, \pr_\mu} \, e^{- \, \int_0^1 \, A_\nu \left(y + t \, q
\right) \, q^{\nu} \, dt}. \eqa this follows from (\ref{exprrr}).
Furthermore, for an arbitrary function $F(p,y)$, we have \bqa {\rm
Tr}_{\mathcal{H}} \int dq \, e^{{\rm i}\, q^\mu \, \pr_\mu} \,
F(q,y) = \int dp \, dy \, dq \, \langle y|e^{{\rm i}\, q^\mu \,
\pr_\mu}|p\rangle \, \langle p|F(q,y)|y \rangle = \nonumber
\\
= \int dp \, dq \, dy \, e^{{\rm i} \, q^\mu \, p_\mu}F(q,y)=\int
dy \, F(0,y). \eqa This proves (\ref{aaa}). Thus: \bqa S_{RR} =
\mu_9 \, \int_{M^{(10)}} [C_{RR}\wedge e^{-\frac{1}{2\pi} \,
F_{(2)}[A]}]_{top}, \eqa which is what we expect for $D9$-branes.
In conclusion, we could equivalently turn on the gauge fields in
$x$-space and in $y$-space.

\subsection{Towards Universal Description of $Dp$-brane RR-couplings}

In this subsection we give a general construction of the RR gauge
field couplings for arbitrary $D$-brane background.

Let us start
with a fixed infinite-dimensional bundle over the space-time with
the fiber identified with the space of sections of the spinor
bundle over the base space times the two-dimensional vector space.
We start with the construction of a superconnection corresponding
to $n_+$ $D7$ branes and $n_-$ of $\overline{D}7$ parallel branes.
Let $(y_a=y_{9},y_8)$ be coordinates in the orthogonal plane and
$W(y_9,y_8)$ is a function defined by the condition that its only
critical points are the positions of $D7$ and $\overline{D}7$
branes in the plane $(y_9,y_8)$ (the sign of the Hessian $\pr^2 W$
defines the sign of the RR charge). Consider now the
superconnection: \bqa \A_{s,t}=d+\frac{1}{\sqrt{t}} \,
\gamma_{\mu}\, \left(x^{\mu} - y^{\mu}\right ) +
\frac{1}{\sqrt{s_1}} \, \Gamma_{a}\, \pr^{a}W(y) +
\frac{1}{\sqrt{s_2}}\, \widehat{\Gamma}_{\mu} \, \widehat{p}^{\mu}
\eqa
In this formula the additional two-dimensional vector space
is interpreted as the representation of the Clifford algebra of
the cotangent space to the transverse space (the notations are in
agreement with (\ref{Clif1}).
Note that this representation is
very close to the expression that enter the Supersymmetric Quantum
Mechanic interpretation of Morse theory \cite{Morse}.\footnote{
One could say that we have $N=1$ SUSY Quantum mechanic in the
directions orthogonal to D-brane and $N=\frac{1}{2}$ SUSY Quantum
mechanic in the directions parallel to D-brane} In fact, the
calculation of the Chern character for this connection in the
limit $t,s_i \rightarrow 0$ gives the expression with the
insertion of the projector in the $y$-integral: \bqa
\det(\pr_{a}\, \pr_{b}^{\phantom{\frac12}}W ) \,
\delta^{(2)}(\pr_{a}^{\phantom{\frac12}}W(y)) \eqa along with
$\delta^{(10)}(x-y)$ and $\delta^{(10)}(\widehat{p})$. Trivial
calculations reduce the full expression to the one given in the
first part of the paper and gives rise to the Chern character form
representing $n_+$ $D7$-branes $n_-$ $\overline{D}7$-branes
($n_{\pm}$ is number of critical points of $W(y)$ with the
positive and negative indexes).

Note that at the intermediate step after taking the limit
$s_1\rightarrow 0$ we get an infinite dimensional bundle with the
fiber naturally identified with the space of sections of the super
bundle of spinors on the auxiliary space of two dimensions lower
multiplied by the finite dimensional super-bundle $E=E_+\oplus
E_-$ with ${\rm dim} \,E_{\pm}=n_{\pm}$. Let us represent the
ten-dimensional space-time $\R^{10}$ in the factorized form
$\R^{10}=\R^{2}\otimes \R^{8-p-1}\otimes \R^{p+1}$. One could
identify the bundle $E$ with the bundle of the representations of
Clifford algebra of the cotangent space to $\R^{8-p-1}$ (similar
to identification of the gauge bundle and tangent bundle in string
compactifications). Now we could again use the same localization
procedure. Suppose we would like to get the superconnection
leading to the RR-coupling with $Dp$ parallel branes orthogonal to
$\R^2\otimes \R^{8-p-1}$. Let $W^{(p)}(y)$ be a function on
$\R^{8-p-1}$ whose critical points define the positions of
$Dp$-branes. Then the following superconnection leads to the RR
coupling with $Dp$ branes: \bqa \A_{s,t}=d+\frac{1}{\sqrt{t}} \,
\gamma_{\mu}\, \left(x^{\mu} - y^{\mu}\right ) + (\Gamma_{a}\,
(\pr^{a}(\frac{1}{\sqrt{s_1}}W(y_9,y_8)+
\frac{1}{\sqrt{s_p}}W^{(p)}(y_7,\cdots,y_{p+2})))) + \\ \nn
+\frac{1}{\sqrt{s_2}}\, \widehat{\Gamma}_{\mu} \,
\widehat{p}^{\mu} \eqa Here the summation is over $a=9,8,\cdots
(p+2)$.

Iteration of this procedure leads to the expressions for RR gauge
field couplings discussed at the first part of the paper. The set
functions $W^{(p)}$ emerged in this iterative procedure is rather
special and connected with the special geometry of $D$ brane
configurations. Consideration of arbitrary functions leads to the
expression for RR gauge field coupling with an arbitrary
configuration of $D-\overline{D}$ branes.

\section{Conclusions}

In this note we have proposed a universal description of the
couplings of RR gauge fields with open strings in arbitrary
backgrounds of the $Dp-\overline{Dp}$ for $p<9$ in terms of the
superconnection on some naturally defined infinite-dimensional
bundle. This description is given in terms of the differential
operators in the auxiliary space (which is a kind of a double of
genuine space-time where string theory is defined) depending on
the point of space-time. From the mathematical point of view this
construction is very close to Beilinson construction of the
holomorphic bundles on $\mathbb{P}^n$ \cite{Beil}.

In conclusion let us give  some comments on the interpretation of
this construction within string theory. As was already discussed
in the introduction, open strings (with unspecified boundary
conditions) lead to the differential operators as the crude
approximation of the open string algebra --- only end points of
the open string are taken into account. The real picture is
obviously more difficult and the first correction is that we
should represent an open string state not as an interval but as a
"slice of pizza" part of a closed string world-sheet
\cite{WitDoub,Strominger}. It is defined in the first
approximation by {\em three points} --- two of them are "ends" of
the open string and the third one which is at the "apex" of the
slice is responsible for the interaction with closed string
vertexes. The description we encounter in the last section is very
close to this qualitative picture. Curiously enough the very rough
approximation of the open string algebra leads to a satisfactory
description of the couplings of the open strings to RR gauge
fields. Let us note at the end that the considerations of this
paper, though inspired by string field theory, were rather formal.
This allows us to present the logic behind the construction in the
most explicit way. The detailed comparison/derivation of these
results in terms of the first quantized string theory will be
given elsewhere. It is also interesting to construct the natural ``geometric"
action for the algebra of differential operators in the lines of
\cite{AFS}.

{\bf Acknowledgements:} We are grateful to M. Douglas, Hong Liu,
P. Horava, G. Moore, A. Morozov, N. Nekrasov and Y. Oz for useful
discussions. E. T. A. is grateful to A. Losev for interesting
discussions. S. L. Sh. would like to thank Rutgers New High Energy
Theory Center for hospitality. The work of E. T. A. was supported
by a NATO Science Fellowship and by RFBR 01-01-00548. The work of
A. A. G. was partially supported  by Grant for Support of
Scientific Schools 00-15-96557 and by RFBR 01-01-00548. The work
of S. L. Sh.is supported by OJI award from DOE.

\end{document}